\documentclass[reprint,aps,preprintnumbers,amsmath,amssymb]{revtex4-2}

\usepackage[colorlinks=true,
            linkcolor=purple,
            citecolor=blue,
            urlcolor=blue]{hyperref}

\usepackage{graphicx}
\usepackage{dcolumn}
\usepackage{bm,xcolor}

\begin{document}

\preprint{IPPP/26/45}

\title{Discovering the Axiverse via Fifth Forces}

\author{Martin Bauer, Francesca Chadha-Day, and Alexander Eberhart}
 \affiliation{Institute for Particle Physics Phenomenology, Department of Physics, Durham University, Durham, DH1 3LE, United Kingdom}

\begin{abstract}
If the Universe is described by string theory or other extensions of the Standard Model, several light, axion-like states are predicted to exist. However, most experiments only search for one of these states. We show that searches for fifth forces can't assume a non-relativistic potential induced by a single axion field. We discuss the strength of the non-relativistic spin-dependent and spin-independent potentials in the most general case of $N_a$ different axions and show how the shape of the potentials can be used to distinguish between different axiverse scenarios.
\end{abstract}

\maketitle

\section{Introduction}
\label{sec:intro}
Extensions of the Standard Model of particle physics (SM) are notoriously hard to test experimentally if the new physics scale is high compared to the electroweak scale. This is the fate of ultraviolet completions that attempt a unified description of the SM and general relativity at the Planck scale. If new physics is realised at the Planck scale $M_\text{Pl}\approx 10^{19} \, \mathrm{GeV}$, any effective interaction between SM particles must be suppressed by that scale, and producing the corresponding states would require energies far above the capacities of accelerators. An exception to this rule is pseudo-Nambu-Goldstone bosons or \emph{axions}, predicted in any theory in which approximate global symmetries are spontaneously broken. UV completions typically have many global symmetries that are broken in the SM, both internal, e.g. flavour symmetries, or space-time symmetries, e.g. extra dimensions. In string theory, the number of axions is related to the explicit compactification scenario and can range from a few to $10^5$ axions~\cite{Douglas:2006es, Arvanitaki:2009fg}. 
The prediction of several axion states in string theory has been called the ``axiverse''. However, many experimental searches for axions focus on detecting a single, light state. The sensitivities of such experiments to several light axions are non-trivial \cite{2107.12813,2311.13658,2512.16837}.

We propose to use searches for fifth forces to discover the axiverse. The pseudoscalar structure of axion interactions with fermions 
\begin{align}\label{eq:Lagdell}
\mathcal{L}_\psi =\frac{1}{2}(\partial a)^2-\frac{m_a^2}{2}a^2+\sum_\psi\frac{\kappa_\psi}{2}\frac{\partial_\mu a}{f}\bar \psi \gamma^\mu \gamma^5 \psi\,, 
\end{align}
induces a spin-dependent potential in the non-relativistic limit. All axions with masses $m_a<1/r$ contribute to the resulting fifth force~\cite{Moody:1984ba, Dobrescu:2006au}. In the case where these contributions add constructively (as they necessarily do for a single fermion species), the resulting fifth force is enhanced by the number of light axions, $N_a$.  In contrast, the leading contribution to the force between unpolarised targets induced by axions is a one-loop effect and results from the exchange of pairs of axions with a potential scaling like $V(r)\sim 1/r^3$~\cite{Bauer:2023czj}. Even though this potential decays fast, the experimental sensitivity, e.g. in torsion balance searches~\cite{Adelberger:2006dh}, can provide constraints competitive with searches for dipole-dipole interactions with spin-polarised $^3$He and K atoms \cite{Vasilakis:2008yn} and $^{87}$Rb~\cite{Almasi:2018cob}. In the axiverse, this force can be enhanced by a factor $N_a^2$. 
This enhancement of fifth forces can become significant for large numbers of light states. If the masses of these states differ, the non-relativistic potentials for the fifth forces reflect several length scales at once. It is usually a sum or mixture of different radial profiles rather than one clean single-range force law. In principle, a careful measurement of the spin-dependent and the spin-independent forces could
allow us to count the number of axion states. We calculate the potential for the fifth force induced by the 1-loop exchange of multiple axions, including interference terms and full mass dependence, and compare the projections for fifth force searches with spin-dependent and spin-independent targets. By evaluating the contributions to the spin-independent potential associated with axion mass spectra predicted by different string theories, we can construct effective potentials that can distinguish between these models. 

\section{Fifth Forces in the Axiverse}
We focus on interactions between axions and nuclei, for which we define the operators 
\begin{align}\label{eq:lag}
    \mathcal{L} = \sum_{i=1}^{N_a} \frac{\partial_\mu a_i}{2 f_i} \bar N g_{i} \gamma^\mu \gamma^5 N + \sum_{i,j=1}^{N_a} \frac{a_i a_j}{f_i f_j} \bar N c_{ij} N
\end{align}
where $N=(p,n)$ denotes the proton and neutron fields, and $g=(g_p, g_n)$ and $c=\text{diag}(c_p,c_n)$ parameterise the linear and quadratic axion couplings and $f$ is the axion decay constant. The shift-symmetry breaking interactions are dimensionful and, for example, scale as $c_{ij}\sim \frac{m_\pi^2}{\rm GeV}$, for axions coupling to gluons. For axions that don't interact with gluons, the effective coefficient is typically suppressed by the axion mass~\cite{Bauer:2023czj}.

The force induced by the exchange of axion pairs can be calculated by evaluating the Laplace transform
\begin{align}
    V_2^{(N_a)}(r) = \frac{1}{4 \pi^2 r} \sum_{i,j=1}^{N_a} \int_{(m_i+m_j)^2}^\infty \mathrm{d}t \, \rho_{i j}(t) e^{-\sqrt{t} r} \,,
\end{align}
of the spectral density function $\rho_{i j}$ for the exchange of axions with masses $m_{i}$ and $m_{j}$ between nucleons $N_\alpha$ and $N_\beta$ with nucleon masses $M_\alpha$ and $M_\beta$ and axion couplings $g^\alpha_i$ and $g^\beta_j$, $c^\alpha_{ij}$ and $c^\beta_{ij}$, respectively. 
The spectral density as a function of the four–momentum transfer squared $t$ is given by
\begin{align}
    \rho_{i j}(t) &= \frac{\sqrt{\tau_\delta \tau}}{128 \pi f_i^2 f_j^2} \Bigg\{g^\alpha_i g^\beta_i g^\alpha_j g^\beta_j - 16\frac{c^\alpha_{ij} c^\beta_{ij}}{t} \\\
    &+ \bigg[  \frac{2g^\alpha_{i} g^\alpha_{j} c^\beta_{ij}}{M_{a}} + (\alpha \leftrightarrow \beta) \bigg] \bigg( 1+ \frac{\tilde{t}^2}{\tau_\delta\tau} \bigg)\!\Bigg\}\,,
    &\quad
\end{align}

where $\tau = t - (m_i+m_j)^2$, $\tau_\delta = t - (m_i-m_j)^2$, and $\tilde{t} = t - m_i^2 - m_j^2$. In the special case where all axions have the same mass $m_a$, the spin-independent potential can be evaluated analytically and reads

\begin{align}
    \label{eq:V2_equal_masses}
    V_2^{(N_a)}(r) &= \frac{1}{128 \pi^3} \sum_{i,j=1}^{N_a}\frac{1}{f_i^2 f_j^2}\Bigg\{\frac{3}{r^5}g^\alpha_{i} g^\beta_{i} g^\alpha_{j} g^\beta_{j}\mathcal{K}(x_a)   \notag\\
    &+ \frac{12}{r^5} \bigg[  \frac{g^\alpha_{i} g^\alpha_{j} c^\beta_{ij}}{M_{\alpha}} + (\alpha \leftrightarrow \beta) \bigg]\!\bigg[\mathcal{K}(x_a)+\frac{x_a^4}{48} K_0(x_a) \bigg]\notag \\
    &-\frac{8}{r^3} c^\alpha_{ij} c^\beta_{ij} x_a K_1(x_a)\Bigg\} \,,
\end{align}
where
\begin{align}
\mathcal{K}(x_a)= \frac{x_a^2}{2} K_0(x_a) + \bigg(x_a+\frac{x_a^3}{6}
    \bigg) K_1(x_a) 
\end{align}

and $K_0(x_a)$ and $K_1(x_a)$ are hyperbolic Bessel functions of the second kind and $x_a=2m_a r$.  In the limit $x_a\to 0$, the function $\mathcal{K}(x_a)\to 1$, because $\lim_{x_a\to 0} x_a K_1(x_a)= 1$, while all other terms vanish.

For comparison, we also give the spin-dependent potential for the exchange of $N_a$ axions,

\begin{align}\label{eq:SDpotential}
V_{1}^{(N_a)}(r)
&= -\sum_{i=1}^{N_a} \frac{g^\alpha_{i} g^\beta_{i}}{16\pi f_i^2}e^{-m_i r}\\
&\times\Bigg[
\vec{S}_1\cdot \vec{S}_2
\left(
\frac{m_i}{r^{2}}+\frac{1}{r^{3}}+\frac{4\pi}{3}\delta^{(3)}(\mathbf r)
\right)
\nonumber\\[4pt]
&
-\vec{S}_1\!\cdot\hat r\,\,
\vec{S}_2\!\cdot\hat r
\left(
\frac{m_i^{2}}{r}+\frac{3m_i}{r^{2}}+\frac{3}{r^{3}}
\right)
\Bigg]
\nonumber\\[6pt]
&\approx
-\sum_{i=1}^{N_a} \frac{g^\alpha_{i} g^\beta_{i}}{16\pi f_i^{2} r^{3}}
\left[
\vec{S}_1\cdot \vec{S}_2
-3\,\vec{S}_1\!\cdot\hat r\,\,
\vec{S}_2\!\cdot\hat r
\right] \notag\,,
\end{align}
where the last line is the spin-dependent potential in the limit $m_i\to 0$.
Since the spin-dependent potential is induced by linear axion exchange, there are no cross terms and the cross section for the exchange of $N_a$ light axions with the same nucleon coupling strengths is given by $V_1^{(N_a)}(r) = N_a V_1(r)$. 

For large values of $N_a$, the contributions of higher-order terms in \eqref{eq:lag} and higher loop contributions become as relevant as the leading-order term, because the combinatorical factors are so large that the perturbative expansion breaks down. Comparing the leading-order term for the spin-independent potential with the contribution from the exchange of four axions in each diagram gives rise to a perturbative limit 
\begin{align}
N_a\lesssim 16\pi^2 \,f^2 r^2\,,
\end{align}
assuming the coefficients $g$ are independent of $N_a$.
For $f$ in the GeV range and $r \sim 10^{-4}$, the perturbative limit is $N_a\lesssim 10^{25}$, so we can be confident in our perturbative calculations for all scenarios considered here.

\section{Comparing spin-dependent and spin-independent forces}

Searches for fifth forces are laboratory tests for light degrees of freedom beyond the SM. Experimental searches focus on the effective potential induced by the exchange of single particles such as \eqref{eq:SDpotential} for $N_a=1$ in the case of spin-dependent couplings~\cite{Vasilakis:2008yn, Dong:2025oer}. The effective spin-independent potential from particle-pair exchange is instead modeled using the general form
\begin{align}\label{eq:genpotential}
    V_k(r)=\frac{c_k}{r^k}e^{-m r} \,.
\end{align}
For axion masses $10^{-8} \, {\rm eV} \leq m_a \leq 10^{-2} \, {\rm eV}$, fifth force searches are the most sensitive laboratory searches for axion couplings to nucleons~\cite{OHARE, Adelberger:2006dh,Vasilakis:2008yn, Dong:2025oer}.
In a scenario in which multiple light axions are present, both the spin-dependent and the spin-independent potentials can differ from the form assumed in experimental searches. A constant axion background, as in the case of axion dark matter, can further modify the resulting force~\cite{Grossman:2025cov}. 
If all axions are very light with respect to the length scales probed in the experiment $m_i<1/r$, the forces are enhanced by a factor $N_a$ and $N_a^2$ in the case of the spin-dependent and independent force, respectively. In contrast, searches for axion dark matter sensitive to a specific axion mass, e.g. resonances in haloscopes, can only expect a reduced signal in the case of multiple axions contributing to the relic dark matter density. Even in broadband axion dark-matter searches, the expected signal is altered. Rather than a single axion dark-matter candidate, one should expect multiple weaker signals from the subset of axions within the detector’s search window, modifying the experiment’s sensitivity~\cite{JEDI:2022hxa, Gavilan-Martin:2024nlo, Lee:2022vvb, Bloch:2021vnn}. 
Astrophysical processes, such as neutron-star cooling~\cite{Buschmann:2021juv} and axion emission from SN1987A~\cite{Carenza:2019pxu}, are sensitive to all light degrees of freedom and provide the dominant indirect constraints over the relevant parameter space, although the robustness of the supernova bounds has been questioned in light of uncertainties in modeling the core collapse~\cite{Bar:2019ifz}. An exception are scenarios in which the axions have \emph{only} quadratic interactions with nucleons. In this case the emission rate from supernovae is suppressed~\cite{Bauer:2025hdq}, and the dominant dark matter searches are sensitive only to spin-dependent interactions~\cite{Bloch:2021vnn}. Even though these models aren't the focus here, fifth-force searches provide the best search strategy in this case, in particular when enhanced by a large number of light states by a factor of $N_a^2$. 
If the axion fields are so light that the axion Compton wavelength extends over distances comparable to the Earth’s radius $m_a R_\oplus\lesssim 1$, the Earth acts coherently as a macroscopic source and can generate an appreciable gradient in the local axion fields~\cite{Hook:2017psm, Balkin:2020dsr, Banerjee:2022sqg, Bauer:2024hfv, Bauer:2024yow}. In this regime, tests of the equivalence principle such as MICROSCOPE~\cite{Touboul:2017grn, Touboul:2022yrw} provide a stronger constraint on models with quadratic interactions. For such very light axions, black hole superradiance provides the dominant constraints for axions with masses $10^{-19} \, {\rm eV} \lesssim m_a \lesssim 10^{-18} \, {\rm eV}$ and decay constant $f \gtrsim 10^{14} \, {\rm GeV}$ and with masses $10^{-13} \, {\rm eV} \lesssim m_a \lesssim 10^{-12} \, {\rm eV}$ with decay constant $f \gtrsim 10^{12} \, {\rm GeV}$ \cite{Hoof:2024quk,Caputo:2025oap}. Such searches rely on mass resonance and so are sensitive to each axion mass eigenstate individually. However, black hole superradiance can be seeded by a quantum fluctuation and hence does not depend on the relic density. It can therefore be a highly sensitive probe of scenarios with large numbers of light axions \cite{Mehta:2021pwf}, although current observations are only sensitive to the relatively narrow mass ranges above.

\section{Axiverse Scenarios}

In the following, we discuss the shape of the effective, non-relativistic potential induced by the exchange of multiple, different axions with the interactions~\eqref{eq:lag}. We focus on three examples motivated by UV models of extra-dimensional axion models \cite{hep-ph/9912455,1106.4546,1107.0721,2412.00179} and string theory \cite{hep-th/0605206,Arvanitaki:2009fg} as different, effective models for the axiverse. We discuss and demonstrate the impact of multi-axion exchange in comparison to the potential induced by the exchange of a single axion $N_a=1$.

\subsection{Kaluza Klein ALP}
The first benchmark model is the Kaluza Klein ALP model considered in \cite{2512.16837}, based on the Kaluza Klein axiverse with no instanton contributions \cite{Dienes:1999gw}. The masses and decay constants are given by

\begin{equation}
m_n = \sqrt{m_0^2+\left(n \mu_1\right)^2},
\end{equation}
\begin{equation}
f_n = f_a=\text { const },
\end{equation}

where $m_0 = 10^{-4} \, {\rm eV}$ and $\mu_1 = 5 \times 10^{-5} \, {\rm eV}$.

\subsection{Kaluza Klein Maxion}
We next describe the Kaluza Klein Maxion benchmark model considered in \cite{2512.16837} for a Kaluza Klein axiverse in which the QCD axion parameters are maximally far from the canonical QCD band \cite{2412.00179}. The masses and decay constants are given by

\begin{equation}
m_n = \left(n-\frac{1}{2}\right) \mu_1,
\end{equation}
\begin{equation}
f_n = f_a  \frac{\sqrt{2}}{2 n-1}, 
\end{equation}

where $\mu_1 = 5 \times 10^{-5} \, {\rm eV}$ is the mass of the lightest KK mode. In order to prevent the decay constants from becoming arbitrarily small for increasing numbers of total axions $N_a$, we choose to rescale the $f_n$ as
\begin{align}
    f_n \to f_n^{(N_a)} = f_a \frac{2 N_a-1}{2 n-1} \,,
\end{align}
meaning for any number of axions $N_a$, the heaviest axion has the decay constant $f_{N_a}^{(N_a)} = f_a$. This allows us to isolate the effect of increasing the number of axions, rather than the effect of decreasing the decay constants.

\begin{figure*}[t!]
    \centering
    \includegraphics[width=.5\linewidth]{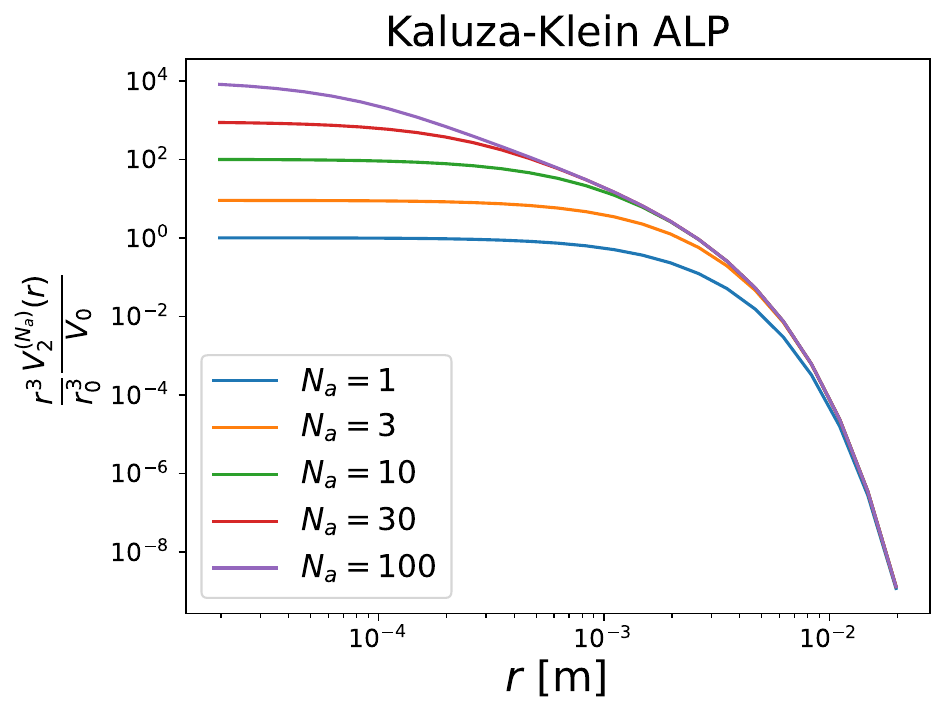}\includegraphics[width=.5\linewidth]{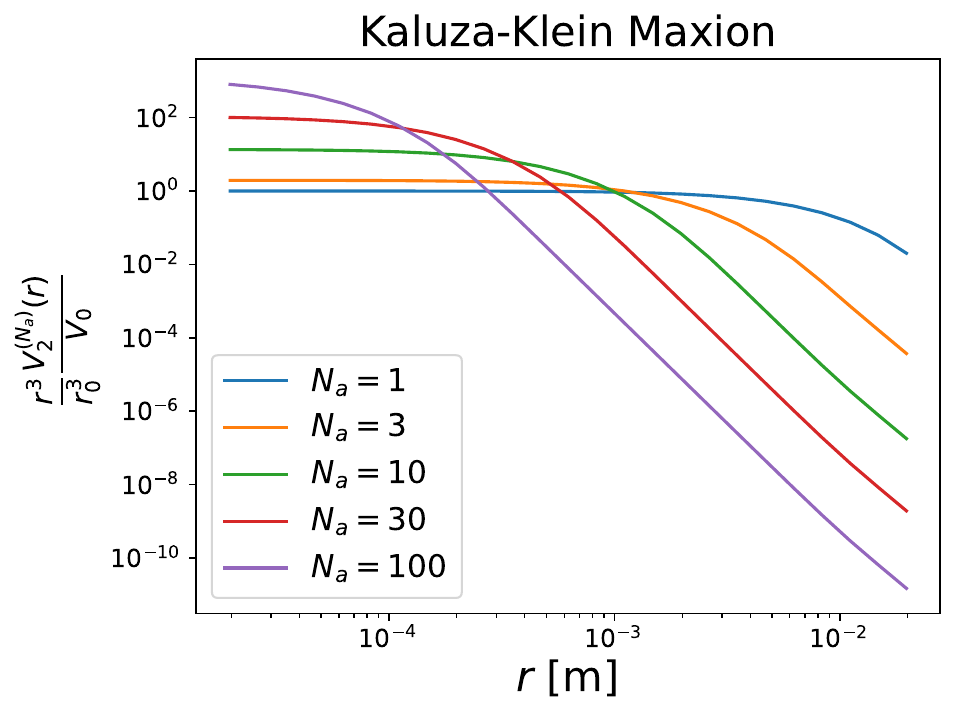}
    \caption{Potentials $V_2^{(N_a)}$ generated by pair exchange of $N_a$ Kaluza-Klein ALPs (left) or Kaluza-Klein Maxions (right) over the distance $r$, normalised such that the $y$-value is $1$ for $N_a=1$ at small distances. The normalisation factors are $r_0 = 100.0 \,\mathrm{eV}{^{-1}} = 1.97 \times 10^{-5}\,\mathrm{m}$ and $V_0 = -1.77 \times 10^{-59}\,\mathrm{eV} \times (10^8 \, \mathrm{GeV} / f_a)^4$. The potentials are calculated with $g_{i} \sim 1$ and $c_{ij} \sim M_n$.}
    \label{fig:V2_KKboth_norm}
\end{figure*}

\subsection{Type IIB String Theory}

The number of axions in type IIB string theory is equal to the Hodge number $h^{1,1}$ of the compactification Calabi-Yau. The Kreuzer-Skarke database \cite{hep-th/0002240} allows construction of toy models with $h^{1,1} \leq 491$ \cite{2309.13145}.

A scaling approximation for the masses and decay constants of axions in type IIB string theory is found in \cite{2507.12516}, where it is shown that the axion effective field theory can be approximated using only data on the compactification manifold's divisor volumes. In particular, \cite{2507.12516} consider the normalised divisor volumes

\begin{equation}
\hat{\tau}^a:=\frac{\log_{10} \left(\tau^a\right)}{\langle\log_{10} (\vec{\tau})\rangle_X},
\end{equation}

where $\tau^a$ are the $h^{1,1} + 4$ prime toric divisor volumes of the compactification Calabi-Yau threefold $X$, normalised such that the smallest volume is  $\hat{\tau}^a_{\rm min} := 1$. The denominator is the average of all such divisor volumes in $X$. The normalised divisor volumes are shown to be approximately distributed as

\begin{equation}
\label{eq:NormalisedDivisorVolumes}
p(\hat{\tau})=\frac{\pi}{2} \hat{\tau} e^{-\frac{\pi \hat{\tau}^2}{4}}.
\end{equation}

The mean divisor volume is approximately given by

\begin{equation}
\label{eq:MeanDivisorVolume}
\langle\log_{10} (\vec{\tau})\rangle=a-b \times\left(\log _{10} h^{1,1}\right)^{-c},
\end{equation}

where the constants $a \simeq 3.433$, $b \simeq 5.404$ and $c \simeq 4.050$ are obtained by fitting to the Kreuzer-Skarke database. Combining \eqref{eq:NormalisedDivisorVolumes} and \eqref{eq:MeanDivisorVolume}, we can draw a spectrum of $h^{1,1}+4$ divisor volumes $\tau^{i}$ typical for a Calabi-Yau threefold of Hodge number $h^{1,1}$. To obtain the corresponding axiverse spectrum, we first approximate the basis of divisors using the smallest $h^{1,1}$ divisors, amounting to discarding the largest four $\tau^i$. The axion masses and decay constants are then given by

\begin{align}
m_i & =\frac{M_{pl}\left(\tau^i\right)^{3 / 4}}{\left(\tau^{\max }\right)^{3 / 4}} e^{-\pi \tau^i}  \\
f_i &=\frac{M_{pl}}{\left(\tau^{\max }\right)^{3 / 4}\left(\tau^i\right)^{1 / 4}},
\end{align}

where $\tau^{\max }$ is the maximum of all $h^{1,1} + 4$ divisor volumes. \\

In this work, we are concerned with the axion-fermion couplings, which, unlike the axion-photon couplings, are not currently calculable within type IIB string theory. In the case of the axion-photon couplings, it was found in \cite{2309.13145} that $g_{a \gamma \gamma}^i \ll 1/f_i$ for the majority of axions. Nevertheless, to make progress in characterising experimental signatures for different Hodge numbers, we will assume $g_{i} \sim 1$, as expected in naive effective field theory, for all three of our benchmark models. The shift symmetry breaking coupling $c_{ij}$ has mass dimension one - the relevant mass scale for this coupling is UV dependent. In particular, in the string axiverse case the light axions would in general not couple significantly to gluons, as such a coupling would give a contribution to their mass via mixing with the pion. We therefore consider shift symmetry breaking couplings in the range $0 \leq c_{ij} \leq M_N$. In the case $c_{ij} = 0$, the only contribution to the spin independent potential comes from the first term in \eqref{eq:V2_equal_masses}, leading to a (highly suppressed) potential scaling as $V_2 \propto r^{-5}$ rather than $V_2 \propto r^{-3}$ when $c_{ij} \neq 0$. The results below are given for the other endpoint of our range $c_{ij} = M_N$, with smaller values leading to a straightforward rescaling of the spin-independent potential.

\begin{figure}[t]
    \centering
    \includegraphics[width=\linewidth]{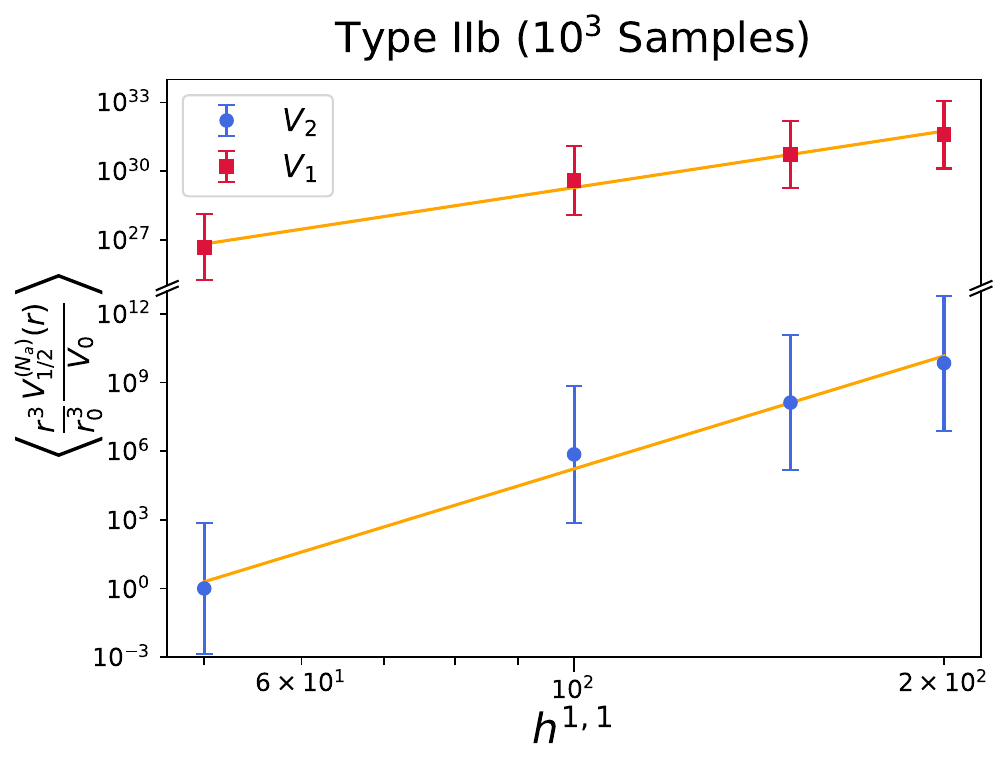}
    \caption{Spin-dependent potentials $V_1$ and independent potentials $V_2$ generated by $h^{1,1}$ axions in type IIb string theory averaged over $10^3$ sampled topologies. The normalisation factors are $r_0 = 100.0 \,\mathrm{eV}^{-1} = 1.97 \times 10^{-5}\,\mathrm{m}$ and $V_0 = -1.05 \times 10^{-78}\,\mathrm{eV}$. The potentials are calculated with $g_{i} \sim 1$ and $c_{ij} \sim M_n$. The best linear fits give slopes of $s_1 = 8.2 \pm 3.2$ for $V_1$ and $s_2 = 16.4 \pm 6.4$ for $V_2$.}
    \label{fig:V1_V2_IIb_mean_1000_samples_fit}
\end{figure}
\section{Results and Discussion}

The potentials for the Kaluza-Klein models as a function of $r$ are shown in Fig.~\ref{fig:V2_KKboth_norm}. In both benchmark scenarios, at short distances the potential follows the $1/r^3$ behaviour typical of a two-particle exchange up to some threshold distance corresponding to the mass of the heaviest axion. As the distance increases further, additional axion contributions ``turn off'' as they are too heavy to mediate a force over the longer distance. For large values of $N_a$, the effective potential increases faster for shorter distances than any model with a single axion could explain. For example, for $N_a=100$ the effective potential scales like $V_2^{(100)}(r)/V_2^{(1)}(r)\sim N_a^2/r$ in the Kaluza-Klein ALP scenario. This characteristic potential shape could be used to indicate the number of axions in the axiverse and their masses.

The number of axions also influences the scale of the potential at small distances, where all axions still contribute to the fifth force. 
For both the Kaluza-Klein ALP and the Kaluza-Klein Maxion, the potential scales with $N_a^2$.
For the Kaluza-Klein Maxion, the potential increases less rapidly with larger $N_a$ relative to the case $N_a=1$. In this scenario, the heaviest axion is set to have a decay constant of $f_a$, and all lighter axions have larger decay constants, thus contributing less towards the total potential than in the case of the Kaluza-Klein ALPs, where $f_n$ is constant.

When calculating the potential for type IIb string theory, there is significant statistical fluctuation in the values of the potential due to the random sampling of the toric divisor volumes. However, since the majority of the axions are extremely light, the potentials show a near-perfect $r^{-3}$ dependence in the range $r \in [10^{-5}\,\mathrm{m},10^{-1}\,\mathrm{m}]$ arising from the dominant quadratic axion coupling, as described by \eqref{eq:V2_equal_masses}, with only minor deviations. Therefore, we average the values of $r^3 V_{1/2}^{(N_a)}(r)$ for the spin-dependent and independent potentials, which are approximately constant inside the aforementioned distance range, over a set of $10^3$ sampled compactification manifolds. These averaged values are presented in Fig. \ref{fig:V1_V2_IIb_mean_1000_samples_fit}, alongside their respective statistical uncertainties, as a function of the Hodge number $h^{1,1}$. We see that $V_1 \propto N_a^8$ while $V_2 \propto N_a^{16}$. This is because, as well as increasing the number of axion exchange diagrams, increasing the Hodge number also typically decreases the axion decay constants \cite{2507.12516}. For the type IIB string axiverse scenario, fifth force potentials therefore depend very sensitively on the Hodge number, but this is typically seen only in the total potential rather than its distance behaviour. This pattern is seen in calculations of the axion decay constant in the F theory axiverse for axion numbers up to $N_a = 181200$ \cite{Fallon:2025lvn} with the distribution of axion masses including many below $10^{-2}\,{\rm eV}$. Such an axiverse with very high Hodge number could therefore generate a much stronger fifth force than expected from a single axion model. Since the spin-independent potential $V_2$ grows more quickly with increasing $h^{1,1}$ than $V_1$, as long as this tendency continues for large values of $h^{1,1}$, the spin-independent potential may be greater than the spin-dependent potential in theories with a very large number of axions. Assuming we can extrapolate the fits in Fig.~\ref{fig:V1_V2_IIb_mean_1000_samples_fit}, the potentials are equal at approximately $h^{1,1} \sim 10^5$.

\begin{figure}
    \centering
    \includegraphics[width=1\linewidth]{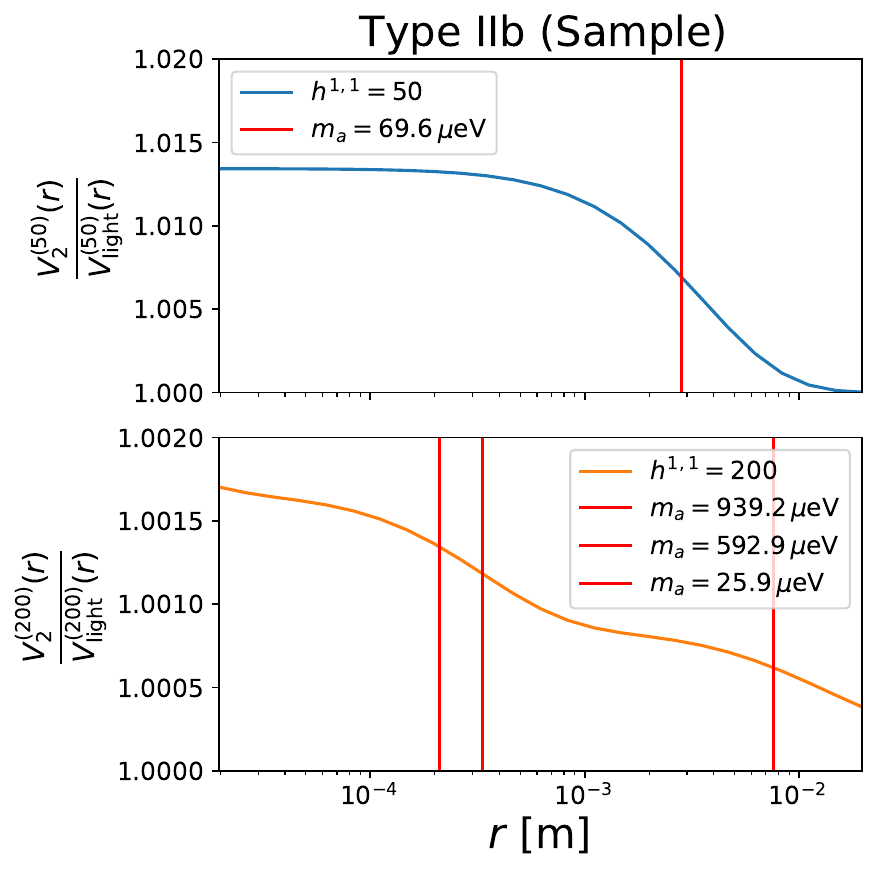}
    \caption{Ratio between the full potential $V_2^{(h^{1,1})}$ and the potential generated by light axions $V_\mathrm{light}^{(h^{1,1})}$ for a sample of $h^{1,1} = 50$ and $h^{1,1} = 200$ axions. The red lines indicate where $m_a r = 1$.}
    \label{fig:V2_IIb_stacked_ratio}
\end{figure}

The main contribution to the total potential comes from axions with $m_a \ll 1/r$. However, the contributions of the higher-mass axions are apparent when factoring out the potential generated by the light axions. This is visualised in Fig. \ref{fig:V2_IIb_stacked_ratio}, where we plot the ratio of the full potential $V_2$ to the potential $V_\mathrm{light}$ generated by all axions with masses $m_a < 10^{-7} \, \mathrm{eV}$ for samples with $h^{1,1} = 50$ and $h^{1,1} = 200$, respectively.

The heavier axions lead to small, step-like changes in the potential that aren't represented by a single potential of the type \eqref{eq:genpotential}. In the top half of Fig. \ref{fig:V2_IIb_stacked_ratio}, we can see that a single axion with a mass of approximately $m_a \approx 7 \times 10^{-5} \, \mathrm{eV}$ leads to an increase in the short-range potential of about one percent. In the bottom half of Fig. \ref{fig:V2_IIb_stacked_ratio}, three axions increase the potential by a combined amount of approximately one permille. While the exact locations and sizes of these steps vary for different samples on account of different masses and different coupling strengths, this general behaviour remains consistent over the different samples. We can thus conclude that, while a larger number of axions greatly enhances the strength of the potential, the relative contributions of any single axion are reduced for a larger number of total axions. For much higher axion numbers as studied in \cite{Fallon:2025lvn}, we may expect to see a modified slope of $V(r)$ as the steps generated by individual heavy axions could not be resolved individually.

\section{Conclusions}
In this work, we have explored the distinctive patterns in the fifth forces generated by different axiverse scenarios. We present a general expression for the non-relativistic potential in extensions of the SM with multiple axion fields. Three benchmark scenarios motivated by Kaluza-Klein models and type IIB String Theory serve to illustrate the effect of multiple axions compared to the experimental benchmark of a single new state. In the case of axion dark matter, fifth force searches can be a more sensitive probe than broadband searches aiming to discover spin-dependent interactions between nucleons and the halo. The non-relativistic potential can be enhanced by the number of light states $N_a$ for the spin-dependent and by $N_a^2$ for the spin-independent case.
Importantly, also the shape of these potentials can differ from any potential that can be modeled by new physics models with only a single new light degree of freedom. We discuss the model dependence of this change in shape of the potential and provide estimates on the uncertainty on generic predictions in the case of string theory axions.

\begin{acknowledgments}
We thank Junyi Cheng and Naomi Gendler for useful discussions. The authors are supported by STFC under ST/X000745/1.
MB acknowledges support from the UKRI Future Leader Fellowship DarkMAP (Ref. no. MR/Y034112/1).
\end{acknowledgments}

\appendix

\nocite{*}
\bibliographystyle{unsrt}
\bibliography{Axiverse}

\appendix

\end{document}